# Electrical transport through self-assembled colloidal nanomaterials and their perspectives

CHRISTIAN KLINKE[1,2(A)]

[1] *Institute of Physical Chemistry, University of Hamburg, Grindelallee 117, 20146 Hamburg, Germany*
[2] *Department of Chemistry, Swansea University, Singleton Park, Swansea SA2 8PP, UK*

**Abstract** – Colloidal nanoparticles developed as interesting objects to establish two- or three-dimensional super-structures with properties not known from conventional bulk materials. Beyond, the properties can be tuned and quantum effects can be exploited. This allows understanding electronic and optoelectronic transport phenomena and developing corresponding devices. The state-of-the-art in this field will be reviewed and possible challenges and prospects will be identified.

**Introduction.** In the recent years, 2D and 3D material printing developed to a serious alternative in product development and even in customized serial production. Beyond pure massive objects it is a vision to be able to process semiconductor channels and metal circuits by easy and cheap assembly techniques, like roll-to-roll processing, microcontact printing, offset printing, dip-coating, and mainly inkjet printing. This could lead to transistors, photo sensors, light-emitting diodes (LEDs), display devices, solar cells, or even whole integrated circuits, such as radio-frequency identification devices (RFID). By 3D printing layered layouts become possible, with insulating but interconnecting layers between the circuits or even 3D device architectures are conceivable. Such a device could include energy harvesting (printed solar cells), energy storage (in batteries or super-capacitors), information processing in a layer in between, and a display technology (LEDs), all processed on flexible substrates. In order to have available corresponding inks inorganic solution-suspended high-quality materials are necessary.

The colloidal synthesis of nanostructures offers the possibility to tune the properties of the materials over a wide range, e.g. the optical bandgap of semiconductor nanoparticles is a matter of size. In solution they can be produced at low material and energy costs, with high flexibility and efficiency in synthesis, and with high tunability in their properties. Conceivable are faster and less expensive processors, more efficient solar cells and fuel cells, batteries and super-capacitors with faster charging periods, more charging cycles, and higher capacities. This is mostly due to the materials' low-dimensional nature which results in quantum effects and/or high surface-to-volume ratios. Their properties might depend not only on their atomic composition but also on their dimensions.

Scientifically, low-dimensional objects represent exceptionally interesting model systems which allow the extension of classical and the development of new concepts with ground-breaking insights [1]. Also for future technologies this opens new perspectives: Examples are single-electron transistors [2,3], well-defined tunable optical emitters [4,5], and efficient charge storage [6,7]. Further they can be used as fluorescence markers in medicine and biology [8,9] in tumor staining [10,11] in hyperthermia [12,13] and as NMR contrast agents [14,15]. In flat panel displays they are used in modern background illumination systems (e.g. in SonyTM displays with TriluminosTM technology) [16]. Thin films of semiconductor nanoparticles can be used as active layers for inexpensive transistors [17,18], photo-detectors [19,20], solar cells [21,22,23], and as chemical sensors [24,25].

**Syntheses, materials and properties.** Inorganic colloidal nanoparticles are usually defined by organic ligands. In most cases, such ligands are long-chained aliphatic hydrocarbons with a functional head group like a thiol, a carboxylic acid, a phosphate, or an amine group. The ligands have different tasks to fulfill: they keep the nanoparticles suspended in solution, preventing nanoparticle aggregation, passivating dangling bonds on the surface of the nanocrystalline particles, and last but not least playing an active role in the definition of size and shape of the nanoparticles.

The use of ligands (also various ones in one synthesis) allows passivation of certain crystal facets. Most of the used ligands (such as deprotonated acids) bind strongly to metal atoms. This is in particular useful in the case of compound semiconductors. In CdSe with hexagonal wurtzite crystal structure ligands such as octadecylphosphonic acid bind selectively to the cadmium sites which allows for a fast growth of the selenium-rich, poorly passivated (000-1) facet [26,27]. This, in turn, results in a rod-like growth of the nanoparticles. The mechanism of organic-inorganic interaction determines the size and shape of a huge number of colloidal nanoparticles. Since the initial steps in the field of semiconducting nanocrystals by Henglein [28], Efros [29], and Brus [30,31], and the introduction of the hot-injection method by Murray, Norris and Bawendi [32], a large variety of shapes has been established. Beside precisely-defined, spherical nanoparticles with tunable diameter [33,34] and rod-shaped ones [35,36], it is possible to synthesize nanoparticles with zig-zag shape [37], cubic geometry [38,39], octahedral form [40], and as wires [41], as platelets [42] and as nanosheets [43]. Furthermore, it is even possible to synthesize core/shell [44,45], dot/rod [46], and tetrapod [47,48] structures. In some cases several shapes can be found in one system by gradually varying the synthesis conditions, e.g. in CdSe [47] and in PbSe [49]. The colloidal synthesis of nanostructured materials has the advantage of being versatile, scalable, and rational in the adjustment of the parameters. The number of materials that can be produced as colloidal nanostructures is almost unlimited. Among them are materials made of pure elements with increasing work function such as Ag, Au [50,51], Pd, and Pt [52] which can be used for catalysis [53,54] or plasmonic devices [55,56]. Less common examples are silicon [57,58], and carbon nanoparticles [59,60]. Further, metal alloy [61,39] and compound semiconductors nanoparticles have been investigated. Typical examples are semiconductors with increasing bulk bandgap like PbSe, PbS, CdSe, CdS, ZnS, and ZnO [62]. Also doped [63] particles and more exotic ones like uranium oxide [64] are feasible. More difficult to produce are ternary materials like $CuGaSe_2$ [65] or $CuInSe_2$ (CIS) [66], $CuInS_2$ [67], and quaternary crystals like $CuIn_{1-x}Ga_xSe_2$ (CIGS) [68] and $CuInS_xSe_{1-x}$ [69,70]. Another effect that can take place, especially for poorly passivated particles, is "oriented attachment": During the later stages of growth nanoparticles agglomerate and merge in order to minimize the total surface (energy) [71,72]. Usually, this happens *via* the most reactive crystal facets of the nanoparticles [73] or by dipole-dipole interaction [74].

[A] E-mail: `klinke@chemie.uni-hamburg.de`

Often nanoparticles suffer from incomplete passivation of the dangling bonds by ligands. This is due to a poor interaction of a facet with the ligands or due to steric hindrance. Poorly passivated semiconducting nanoparticles show low photoluminescence quantum yield and blinking due to the presence of trap states. The passivation can be improved by a post-synthetic ligand exchange for better-passivating ones [75,76]. An even better approach is to grow an inorganic shell around the original particles which has a similar lattice constant but a higher bandgap (e.g. CdSe core with a CdS shell). Depending on the band-alignment both, the optically excited electron and hole are confined in the core (Type I configuration, stabilizing photon emission) or will be separated in the core and the shell (Type II configuration) [77,78]. To improve the optical properties further yet another shell can be grown around the initial core-shell structure (e.g. CdSe/CdS with a further ZnS shell) [79]. This approach was shown to be successful not only for spherical, but also for rod- [80,81] and star-like ones [40].

The electronic confinement of charge carriers in semiconducting nanoparticles has a decisive impact on their optical properties. The most obvious effect is that the absorption and fluorescence wavelength shifts with decreasing nanoparticle volume to shorter wavelengths. In a simple approach this can be explained by the quantum mechanical textbook model of a particle in a box, including the Coulomb interaction of the electron and hole (the exciton) in confined space [82]. Of course the actual optical properties depend further on the solvent, the type of stabilizing ligand, *etc*. Anyhow, this model has been applied successfully in many cases [83,84].

Metal nanoparticles show also interesting optoelectronic effects. Upon absorption of photons collective superficial electron motions can be excited in small metal structures, called plasmons. In defined volumes this leads to spectroscopically clearly distinguishable resonances which are a matter of the size and shape of the structures. In case of gold and silver nanoparticles those absorption bands are in the visible range and lead to colored solutions. Already the ancient Romans produced metallic gold nanoparticles as nicely demonstrated by the famous Lycurgus cup (400 BC, British Museum) [51]. After annealing a mixture of gold salts, soda ash, and sand they obtained transparent, deep red glass. Their method was also used for colored windows in churches of the medieval times like Notre Dame in Paris or the Cologne Cathedral. In 1659, Glauber synthesized colloidal gold in aqueous solution by reduction of gold salts with tin chloride. A similar method was described by Cassius in 1685 in his book "De Auro" [85]. Later on in 1857, Faraday delivered a synthesis and a scientific discussion of colloidal gold which was based on the reduction of gold chloride with phosphorous. These water based colloidal solutions showed red color, too. Nowadays, we attribute the color to the light absorption by the collective surface electron excitation, the so-called plasmon resonance [86,87].

**Assembly of nanoparticles.** Films of nanoparticles can serve as conduction channel for inexpensive transistors [88] or as absorption layer in solar cells [89]. A simple method to produce such films is spin-coating or dip-coating. For more precisely defined structures the well-known Langmuir-Blodgett technique can be exploited [90,91,92]. Usually, it is used to establish monolayers of amphiphilic molecules, but it can also be used to produce well-ordered monolayers of colloidal nanoparticles [93,94]. Also simple scooping methods can result in high quality supercrystalline materials [95]. By this interesting method it is possible to not only deposit thin films as random agglomerates but to establish monolayers with super-crystalline long-range order. Monolayered planar-squared and hexagonal honeycomb single-crystal nanoparticle patterns have been achieved by solvent-evaporation at the air-liquid interface which eventually led to oriented attachment between the nanoparticles [96,97]. Simulations on the bandstructure of 2D honeycomb superstructures show combined features of semiconductors and Dirac bands [98]. Due to the great monodispersity of colloidal nanoparticles, it is possible to create more complex 2D assemblies by self-organization of mixed nanoparticles of various materials, sizes, or shapes. In this way, it is possible to establish superlattices with cubic, hexagonal, tetragonal, and orthorhombic symmetries [99,100]. Those approaches to establish well-ordered thin nanoparticle films have further been exploited for to create electronic devices on a wafer scale [101] or even on larger flexible substrates [102]. This allows creating a new class of materials with emergent physical properties which do not exists in bulk materials and which are tunable by the particle size and shape, and the ligands of the nanocrystals [103]. Precipitating spherical nanoparticles to three-dimensional structures can lead to interesting polymorphic super-crystals such as amorphous, body-centered-cubic (bcc), and face-centered-cubic (fcc) phases [104]. The established phase depends on the inter-nanoparticle distance (determined by the ligand length and the pressure) and the driving force resulting from entropic maximization, from the reduction of the total free energy, and from multiple interactions between the ligand molecules. In nanoparticles with pronounced facets, such as in octahedra, van-der-Waals forces play a prominent role as well and they arrange with vertex-to-vertex and face-to-face [105].

**Electrical transport.** Due to their size of a few nanometers nanoparticles possess a very small electrical capacity and thus, a comparatively large Coulomb energy of $E_C = e^2/C$, with the elementary charge $e$ and the capacity of the nanoparticle $C$. Charging the particles with an electron requires the energy of $E_C/2$. Thus, their size has a decisive impact on their electrical properties and electrons can only be added in discrete steps. This classical effect is called Coulomb blockade. Each further charging of the nanoparticle requires again the same energy which results in a "Coulomb energy ladder" with the pitch of $E_C$. In order to investigate the charge transport over a nanoparticle it must be contacted directly. Otherwise the nanoparticles would be part of the electrode and would possess a much larger capacity. This can be accomplished by contacting the nanoparticles *via* tunnel barriers, which are defined by the length of the insulating organic ligands of the nanoparticles. In order to pass a current through a nanoparticle the electrons must tunnel from a first electrode onto the nanoparticle overcoming the Coulomb charging energy and tunnel to a second electrode [106,107,108]. A third electrode located more remotely does not make direct contact to the nanoparticles (tunneling processes are not possible either); but it can shift the Coulomb energy levels of the nanoparticle up or down electrostatically. In this way, the relative number of electrons located on an individual metal island can be counted, leading to a single-electron transistor (SET) [109,110]. The electrical characteristics for devices with a few nanoparticles is more complex, but still they behave like transistors [111,112]. A thorough review on single-electron devices was written by Likharev [2]. In thin films of colloidal metal nanoparticles the electrical characteristics usually do not show clear signs of single charge transport, though the behavior of the I-V curves can be interpreted as reflecting the existence of tunnel-barriers in the film and Coulomb charging energy of the single metal particles [113,114,115].

Scanning tunneling microscopy (STM) allows identifying individual nanoparticles and the topography of supercrystalline thin films of nanoparticles. The derivative of the I-V curves $dI/dV$ at a fixed spot yields the differential conductance as a function of energy. In case of individual metal nanoparticles it shows directly the Coulomb energy gap and higher conductance level forming the typical Coulomb staircase [116]. In contrast, in thin films of nanoparticles these features smear out, though the Coulomb gap is still recognizable. Surprisingly, it is found that in larger supercrystals some nanoparticles show the features of individual nanoparticles again. They are identified as defects in the periodic structure [117].

The electrical transport through metal nanoparticle assemblies is usually based on either electronic tunneling or thermally excited hopping processes. In order to study the mechanisms the thermal hopping from nanoparticle to nanoparticles due to highly excited electrons (described by the Fermi function) can be switched on at higher and switched off at cryogenic temperatures. Tunnel processes give rise to large resistances due to the gaps between the nanoparticles which need to be crossed. The electrical transport through thin layers of nanoparticles is a function of the distance between the nanoparticles which can be tuned by the ligand length, and the size of the nanoparticles which determines the Coulomb energy gap [118]. At low temperatures and moderate bias voltages the electrons do not possess enough energy to overcome the Coulomb energy $E_C/2$. Thus, the conductance is zero. Higher bias voltages lower the Coulomb levels and electrons can tunnel from particle to particle through the whole film. Thus, at low temperatures the measured current is mainly a tunnel current. At higher temperatures thermally activated electrons contribute increasingly to the current signal and at room temperature usually the thermally activated current is dominant leading to Ohmic behavior [93,119]. The magnitude of the conductance is a matter of particle size (proportional to the single-particle capacity) and interparticle distance [119]. Also the size distribution and the dimensionality of the films play a role. In this way, variable-range hopping (VRH) and nearest-neighbour hopping (NNH) can be distinguished. Whereas the first accounts for relatively unordered films where the transport occurs between sites which are similar in energy but not necessarily adjacent; latter rules mainly highly-ordered assemblies. Since VRH occurs in states around the Fermi level, two different conditions can be present concerning the temperature dependence of the resistivity. With a constant density-of-states (DOS) around the Fermi level Mott (M-) VRH [120] can be determined. Efros-Shklovskii (ES-) VRH [121] is dominant at lower temperatures or in higher disordered films. In this case, the DOS is reduced and a Coulomb gap is opening. M- to ES-VRH [122,123] and NNH to ES-VRH [124] transitions can be observed as a function of temperature and DOS.

As mentioned above, the distance between the particles (tunnel barrier width) is determined by the length of their organic ligands. Thermal annealing of the films can transform a small fraction of the ligands into amorphous, conductive carbon [125]. Annealing time and temperature determine the amount of carbon between the particles, bridging the tunnel barriers. As a result, the conductance through the films increases. This tunability of the film conductance is useful e.g. for sensors: The trade-off between base current and sensitivity can be optimized for the corresponding application.

The position of the Coulomb energy levels can be manipulated by an external electrical field. This holds good not only for individual particles but also for films – at least for monolayers. The fixed interparticle distance in monolayers of metal nanoparticles of about 1 nm allows for an overlap of the wave functions to enable tunneling. Due to the film height of only one monolayer of nanocrystals, screening is limited. In multilayered films the metallic particles would screen the electrical field. If a dielectric layer is deposited onto the metal nanoparticle monolayer a third, gate electrode can be defined on top of it by electron-beam lithography. The application of a gate voltage can then modulate the current through the film due to a shift in the Coulomb energy levels of the individual nanoparticles leading to transistor-type characteristics [126]. The Coulomb charging energy of the individual nanoparticles adopts a similar function like the semiconductor bandgap. Since the charging energy depends on the size of the nanoparticles, a reduction of the diameter and further device engineering can lead to higher working temperatures up to room temperature and larger on/off ratios [127,128]. Due to the involved Coulomb energy ladder, these new types of transistors possess sinusoidal transfer characteristics which are tunable by the size of the particles, the interparticle distances, the used ligands, and the device architecture.

Thin films of metal nanoparticles can also be used as highly sensitive chemical and mechanical detectors due to the exponential dependency of the tunneling process on the distance between the nanoparticles. Such films, which are often based on gold nanoparticles, can be prepared on flexible substrates [129] or even as freestanding layers [130]. The absorption and adsorption of chemical components can cause a swelling of the films and thus, an increase of the tunnel distances or a change of the dielectric environment of the nanoparticles. Both effects usually counteract each other, which can be measured by electrical transport measurements giving a (change in) resistance. Nesting of molecule in the matrix of ligands between the nanoparticles increases the resistance and most volatile molecules increase the effective dielectric permittivity of the nanoparticles' environment and with this the capacity of the nanoparticles. In turn, latter reduces the effective Coulomb charging energy which is reflected in the activation energy [131]. Thus, the size, the function groups, and the dielectric contribution of the analyte give a fingerprint in the electric response of the metal nanoparticles films allowing for chemical detection. The strong dependency on the interparticle distance is also used in electrical strain and pressure sensors [132].

In thin semiconducting films the ligands which define the nanoparticles are disadvantageous since they decrease the electrical performance of the devices. In a landmark paper, Talapin and Murray demonstrated that the replacement of the original ligands by short ones, like hydrazine, can increase the transport properties of nanoparticle transistors enormously [1]. An alternative to such toxic and corrosive ligands is to use thioltetrazole ligands in the synthesis [68]. During thermal annealing, those molecules decompose mainly to smaller, gaseous fragments, yielding virtually ligand-free nanoparticles. The nanoparticles move closer together and the conductance of the films increases. Another approach is to replace the original long-chained aliphatic ligands by short inorganic ligands [88,133], by semiconducting ones [134], or by charged, atomic ligands such as halides [135] or chalcogenides [88]. Goal is to establish band-like transport conditions for high-performance devices [136,137]. In terms of photoconductivity an interesting alternative is to introduce small amounts of metals as cluster-like deposits on semiconducting nanoparticles [94]. In metal-free thin films the photo-excited charge carriers recombine on the nanoparticles due to large tunnel-barriers between the nanoparticles. With a larger amount of metal the charge carriers recombine due to quenching. An optimized amount of metal bridges the semiconductor nanoparticles leading to a maximum on/off ratio in photoconductivity.

Eventually, engineering nanoparticle devices leads to tunable majority charge carrier concentrations, mobilities, and trap densities. The semiconductor base material allows choosing the bandgap to be in the infrared (e.g. with PbS), the visible (e.g. with CdSe), or in the ultraviolet range (e.g. with ZnO), according to the targeted application as transistor, photo-detector or solar cell. The aim is to produce low-cost electronic devices probably mounted on flexible substrates. This has been achieved e.g. for transistors on Kapton foil [138] and even for integrated circuits [17] and solar cells [139].

**Perspectives.** We observe a highly dynamic development in material syntheses and concepts based on colloidal nanostructures which open new areas, beginning from conductive elements, switching units (transistors), LEDs, to solar cells and integrated circuits. Great opportunities for these concepts is provided by low-cost technologies, like functional printing (ink-jet, super-ink-jet, aerosol-jet, spray-painting). Functional printing of (opto-)electronic devices makes the whole electronics industry much more flexible and much cheaper. All kinds of established integrated circuits can be engineered base on nanoparticle devices. This includes e.g. very cheaply printed RFID-like devices for the internet-of-things and other devices which first do not require high performance; though this might come as well. A wealth of applications is conceivable. There are no boundaries for new device concepts. Nowadays, the synthesis of colloidal nanostructures is very versatile. They can be based on a large variety of materials; they span a large range of sizes and shapes, and they can be combined to more complex structures. The properties of the individual nanostructures are tunable (e.g. by their size) and quantum effects can be maintained. This gives them an enormous variety of properties. Researchers and engineers are aiming for inorganic, solution-processable "inks" for circuits with (opto-)electronic properties which can be printed on a lot of types of substrates including flexible ones. The inks can include semiconductors and metals for thin films of resistors, diodes, transistors, circuit paths, capacitors, photo-detectors, chemical and mechanical sensors, solar cells, and light-emitting devices. In terms of LEDs this could lead to entirely new illumination concepts, and active LED-based displays, even as flexible flat panels [140,141]. It becomes apparent that Moore's law gets more and more difficult to be fulfilled. Thus, researchers and engineers think about a redefinition of the prediction: Not the number of transistor or the clock frequency should be in focus, but the user value, as M. Mitchell Waldrop predicts [142]. Thus, there might be a bright future for customizable, flexible, inexpensive devices based on nanoparticles.

\*\*\*

The work of the author is financially supported by the European Research Council *via* an ERC Starting Grant (Project: 2D-SYNETRA (304980), FP7), by the German Research Foundation DFG *via* the project KL 1453/9-2 and by the DFG Cluster of Excellence "Center of ultrafast imaging CUI".